\documentstyle{article}
\begin{document}
\openup6pt

\title {Curious interior solutions of general relativity.}
\author{S.M.KOZYREV \thanks{Email address: Sergey@tnpko.ru}  \\
Scientific center gravity wave studies ''Dulkyn'', \\
PB 595, Kazan, 420111, Russia ,Kazan, Russian Federation \\}
\date{}
\maketitle

\begin{abstract}
In this article, we provide a discussion on a composite class of
exact static spherically symmetric vacuum solutions of Einstein's
equations. We construct the composite solution of Einstein field
equation by match the interior vacuum metric in Schwarzschild
original gauge, to the exterior vacuum metric in isotropic gauge,
at a junction surface. This approach allows us to associate
rigorously with both gauges as a same "space", which is a unique
differentiable manifold M$^4$.

\end{abstract}

\section{Introduction}
The field equations in Einstein gravity theory are non linear in
nature. For a classical field, the differential equations consist
of purely geometric requirements imposed by the idea that space
and time can be represented by a Riemannian (Lorentzian) manifold,
together with the description of the interaction of matter and
gravitation contained in Einstein's equations
\begin{equation}
G_{a b }=T_{a b },  \label{eq4}
\end{equation}
This equation involved a match between a purely geometrical object
so called Einstein tensor \textit{G}, and an object which depends
on the properties of matter the energy-momentum tensor \textit{T}
which contains quantities like the ordinary density and pressure
of matter. Hence, the geometry of 4D spacetime is governed by the
matter it contains. However, this split is artificial. According
to the standard textbooks the general relativity exhibits general
covariance: its laws-and further laws formulated within the
general relativistic framework-take on the same form in all
coordinate systems \cite{Wald}. On the other hand Einstein's
equations (\ref{eq4}) determine the solution of a given physical
problem up to four arbitrary functions, i.e., up to a choice of
gauge transformations. This theory for definition of concept of
co-ordinate system use geometrical terms; meanwhile, the
geometrically interpreted co-ordinate system can emerge here only
together with the geometry, i.e. with definition of metric tensor
g$_{a b }$ \cite{Temchin}. The variables \textit{x}, used in
Einstein's equations, represent co-ordinates of points of abstract
four-dimensional manifold M$^4$ over which there are a set pseudo
Rimanien spaces V$^4$(g), generated by set of solutions g$_{a b
}$. Co-ordinates in each of such spaces have the specific
properties differing from their properties in other spaces
\cite{Gullstrand}. Moreover, a new class of co-ordinates each time
is postulated subsystem
\begin{eqnarray}
C(\mu) g_{a b } &=&0. \label{eq3}
\end{eqnarray}
where C($\mu$)  - some algebraic or differential operators.
Specifically four of ten field equations will not be transformed
by those or other rules, but simply replaced by hand with the new.
In contrast with general relativity, Newtonian theory has as the
geometrical foundation the Euclidean space and absolute time.
Compared with Newtonian gravity, general relativity has one more
independent variable, 9 more dependent ones, the result of these
changes being that the general form of the field equations expands
to 10 partial differential equations (in terms of the metric and
coordinates). Usually used co-ordinate conditions can be wrote in
the form of four equations (\ref{eq3}). Thereby for any four of
components g$_{a b }$ emerge the relations with remaining six and,
probably, any others, known functions. Certainly, equations
(\ref{eq3}) cannot be covariant for the arbitrary transformations
of independent variables, and similarly should not contradict
Einstein's equations or to be their consequence.

The choice of a reference frame in a general relativity quite
often compare to gauges of potentials in an electrodynamics. But
this analogy is the most superficial: this or that gauge is a
problem of exclusively convenience, its this or that expedient
does not influence in any way on a values of physical quantities
and it is not related to observation requirements, - whereas the
choice of co-ordinate system is related to all it essentially.

The specification of the energy-momentum tensor played a very
important role. The exact solutions known have all been obtained
by restricting the algebraic structure of the Riemann tensor, by
adding field equations for the matter variables or by imposing
initial and boundary conditions. One surprise for the reader may
lie in the fact that in a certain gauge any metric whatsoever is a
'solution' of (\ref{eq4}) if no restriction is imposed on the
energy-momentum tensor, since  (\ref{eq4}) then becomes just a
definition of \textit{T}$_{a b }$. Since the field equations are
very complicated, to find solutions physicists makes simplifying
assumptions about the left-hand-side or the right-hand-side. Most
popular simplifying assumptions about the right-hand-side of (1)
are that \textit{T}$_{a b }$ represents vacuum. On the other hand,
as is well known simplifying assumptions about the left-hand-side
often comprise static and spherical symmetry. The most commonly
approach employ the assumption that this configurations describes
the gravitational field outside any body with spherically
symmetric mass distribution.  Some of this solutions where
discovered at early stage of development of general relativity,
but up to now they are often considered as equivalent
representation of some "unique" solution. However, the physical
and the geometrical meaning of the radial coordinate \textit{r}
are not defined by symmetry reasons and are unknown a priori
\cite{Eddington},\cite{Fiziev}. The review by Fiziev outlines that
various vacuum spherical solution in different gauges leads to
existence infinitely many different static solutions of Einstein
equations (\ref{eq4}) with spherical symmetry, a point
singularity, placed at the center of symmetry, and vacuum outside
this singularity, and with the same Keplerian mass \textit{M}.
This paper will present a new feature of well known solution to
the spherically symmetric time independent Einstein system of
equations that govern the behavior of the space-time in the
"interior" vacuum Schwarzschild solution. At the outer boundary
the solutions will be matched to the external vacuum solution, in
a different gauge, for the field equations i.e. the solution in
isotropic coordinates.

\section{Composite solution.}

The metric satisfies to those or other co-ordinate conditions if
some of quantities g$_{a b }$ are linked by some relations, -
whether it be in any point, on a surface or in four-dimensional
domain $\Omega \subset$ V$^4$(g). By definition all co-ordinate
systems in manifold M$^4$ at least locally are equivalent; on the
other hand if in M$^4$ the metric is introduced, properties of
functions g$_{a b }$ in different co-ordinates become different.
Let us consider two distinct manifolds M$^4$$^+$ and M$^4$$^-$.
The metric in these manifolds generated by set of solutions of
field equations (\ref{eq3})  given by g$_{a b }^+$(
\textit{x}$^a$$_+$) and g$_{a b }^-$( \textit{x}$^a$$_-$), in
terms of independently defined coordinate systems
\textit{x}$^a$$_+$ and \textit{x}$^a$$_-$. The manifolds glued at
the boundary hypersurfaces $\Sigma$ $_+$ and $\Sigma$ $_-$  using
independently defining co-ordinates systems \textit{x}$^a$$_\pm$.
A common manifold M$^4$ is obtained by assuming the continuity of
four-dimensional coordinates \textit{x}$^a$$_\pm$ across $\Sigma$,
then g$_{a b }^+$ = g$_{a b }^-$ is required, which together with
the continuous derivatives of the metric components $\partial$
g$_{a b }$ /$\partial$\textit{x}$^c$  $\mid_+$ = $\partial$ g$_{a
b }$ /$\partial$\textit{x}$^c$ $\mid _-$, provide the Lichnerowicz
conditions \cite{Lichnerowicz}.

The resulting manifold $M$ is geodesically complete and possesses
two regions connected by a hypersurface $\Sigma$. Since the
interior vacuum solution is to be matched with an exterior
Schwarzschild solution at the junction surface $r = a$ we use the
Darmois-Israel formalism \cite{Israel}. Using the field equations,
the surface stress-energy tensor can be calculated in terms of the
jump in the second fundamental form across $\Sigma$. Because $M$
is piecewise vacuum solution, the Einstein tensor is zero
everywhere the stress-energy tensor localized at the junction
surface can be calculated
\begin{eqnarray}
T_{a b }=-\delta(\eta) ([K^a_b]-\delta^a_b [K]). \label{eq77}
\end{eqnarray}
The extrinsic curvature, or the second fundamental form, is
defined as
\begin{eqnarray}
K^a_{b \pm} =\frac{1}{2}g^{a b}\frac{\partial g_{b a}}{\partial
\eta} \mid _{\eta=\pm 0} \nonumber
\end{eqnarray}
where $\eta$ the proper distance away from the $\Sigma$ and $[K]$
denotes the trace of $[K_{a b}]= K_{a b}^+ -K_{a b}^-$.

The approach to be taken here to the static spherically symmetric
relativistic configurations involved a match between solutions of
Einstein equations in Schwarzschild original coordinates
\cite{Schwarzschild}, with solutions in isotropic gauge
\cite{Misner} The fields equations in this case simply state the
metric field is just a field in a spherically symmetric space
time. The solution will be given in terms of explicit closed-form
functions of the radial coordinate for the three metric
coefficients. According to the widespread common opinion, the most
common form of line element of a spherically symmetric spacetime
in comoving coordinates can be written as
\begin{eqnarray}
ds^2=-g_{t t } (r, t) dt^2+  g_{r r }(r, t) dr^2 + 2g_{r t }(r,
t)dr dt + \rho(r, t) ^2 (d\theta^2 + \sin^2(\theta) d \varphi^2).
\label{eq7}
\end{eqnarray}
We are free to reset our clocks by defining a new time coordinate
\begin{eqnarray}
t=t'+ f(r). \nonumber
\end{eqnarray}
with \textit{f(r)} an arbitrary function of \textit{r}. This
allows us to eliminate the off-diagonal element g$_{r t }$.
Therefore we shall consider the matching of two static and
spherically symmetric spacetimes given by the following line
elements

\begin{eqnarray}
ds_{\pm}^2=-g_{t t } (r, t)_{\pm} dt^2+  g_{r r }(r, t)_{\pm} dr^2
+ \rho(r, t)_{\pm} ^2 (d\theta^2 + \sin^2(\theta) d \varphi^2).
\label{eq5}
\end{eqnarray}
of M$^4 _\pm$, respectively, where g$_{t t }(r, t)$, g$_{r r }(r,
t)$ and $\rho(r, t)$  are of class C$^2$.

 In his pioneering article Schwarzschild has used
a radial variable \textit{r} and the gauge
\begin{equation}
\det \|g_{a b }\| = 1 ,  \label{eq6}
\end{equation}
for the spherically symmetric static metric. He had fixed the
three unknown functions
\begin{equation}
g_{t t } (r) = 1 - \frac{2M}{\rho(r)}>0,  g_{r r } (r)=-\frac{1}{g_{t t }}<0 \label{e4}\\
\end{equation}
obtaining
\begin{equation}
\rho(r)=\sqrt[2]{r^3 + \rho^3_G },\nonumber
\end{equation}
Another widespread form of the Schwarzschild's solution, was
reached using isotropic gauge (for example in \cite{Misne}),
\begin{eqnarray}
ds^2=-\widehat{g}_{t t } (r)dt^2+  \widehat{g}_{r r }(r) (dr^2+
r^2 d\theta^2 + r^2 \sin^2(\theta) d \varphi^2),\label{eq111}
\end{eqnarray}
 Obviously, these co-ordinates may be used as the same co-ordinate for the metric field with various properties in different domain. To begin with, we consider the solutions of field equations for line elements (\ref{eq111})and  (\ref{eq6}), (\ref{e4}). The result is
\begin{eqnarray}
\widehat{g}_{t t }= \frac{\widehat{\alpha}(1-4 r \widehat{\beta})^4}{r^4} ,\nonumber \\
 \widehat{g}_{r r}=\frac{\widehat{\gamma}(1+4 r \widehat{\beta})^2}{(1-4 r \widehat{\beta})^2},\label{eq1111}
\end{eqnarray}
and
\begin{eqnarray}
g_{t t }= \frac{\alpha r ^4}{(r^3+\rho^3_G)[\alpha(r^3+\rho^3_G)^{1/3}-\beta]} ,\nonumber \\
 g_{r r}=\alpha-\frac{\beta}{(r^3+\rho^3_G)^{1/3}},\label{eq1112}
\end{eqnarray}
where $\alpha, \beta,\widehat{ \alpha},  \widehat{\beta},
\widehat{\gamma}$ arbitrary constants.

Analogous as in stellar models this implies that the one of the
solutions we can assume as "interior" but another as "exterior"
vacuum solution. Now in order for these line elements to be
continuous across the junction we impose an explicit definition
for the arbitrary constants in solutions (\ref{eq1111}),
(\ref{eq1112}). The brief computation yields
\begin{eqnarray}
\alpha= \frac{\widehat{\gamma}}{9-4^{\sqrt{5}}} ,\nonumber \\
\beta= \frac{4\widehat{\gamma}a}{5^{5/6}(9-4^{\sqrt{5}})} ,\nonumber \\
\widehat{\alpha}=-\frac{a}{5^{5/12}},\label{eq1113}\\
\widehat{\beta}= \frac{1+\sqrt{5}}{4 a} ,\nonumber \\
\rho_G= a(\sqrt{5}-1)^{1/3} ,\nonumber
\end{eqnarray}
where $a$ junction radius.

In this case the boundary surface entails via the field equations
a jump in second derivations of metric coefficient, but first
derivatives remains. Thus, from the junction conditions, the
"interior" metric parameters can be determined at the boundary
surface in terms of the "exterior" metric parameters. Note that
for this case $K^a_b$ is continuous across $\Sigma$. Hence,
Darmois-Israel junction conditions are fulfilled. Such
construction allows us to associate rigorously with both gauges as
a same "space", which is a unique differentiable manifold M$^4$.

\section{Discussion}
The gravitational field equations define only the metric over
manifold M$^4$, but not its topological property. The same system
of reference in manifold M$^4$ is interpreted as this or that
co-ordinate system in pseudo - Riemannian space V$^4$(g) at a
different selection of equations (\ref{eq3}), giving some minimum
of information about properties of the required metric.
Nevertheless, Einstein's equations have, of course, a
non-enumerable set of the solutions which are not possessing
properties of the necessary metric. Generally it is impossible to
be assured, that always it will be possible to coat V$^4$(g) with
set of co-ordinate neighborhoods of the same class, i.e. featured
by the same operators $C(\mu)$. It does necessary acceptance
enough wide guesses of the nature of operators $C(\mu)$. The above
consideration confirms the conclusion, that space V$^4$ one can
featured by two or more known class of co-ordinates.  The
solutions of Einstein's equations for such constructions one
assume as "interior" and "exterior" one. Thus, the common geometry
emerges from the junction conditions at the boundary surface. It
must be combined with Darmois-Israel junction conditions. We show
that our approach provides a clear way of showing that the
Schwarzschild solution is not a unique static spherically
symmetric solution, providing some incite on how the current form
of Birkhoff's theorem breaks down. All results can be stated for
four dimensional (pseudo) Riemannian manifolds. It should also be
noted that because general relativity is a highly non-linear
theory, it is not always easy to understand what qualitative
features solutions might possess, and here the composite class of
solutions can used an a guide.


\begin{thebibliography}{99}
\bibitem{Wald}  {\small Wald, Robert M. (1984), General Relativity, University of Chicago Press. }

\bibitem{Temchin}  {\small A.N.Temchin, Uravneniia Einshteina Na Mnogoobrazii, Moskow, URSS,1999, (Russian). }

\bibitem{Gullstrand}  {\small A. Gullstrand, Allgemeine Losung des statischen Eink?rperproblem in der Einsteinschen Gravitationstheorie. Ark. for Mat., Astr. o. fysik, Bd.16, N 8,  (1921)}

\bibitem{Eddington}  {\small A. S. Eddington, The mathematical theory of relativity, 2nd ed. Cambridge, University Press, 1930 (repr.1963).)}

\bibitem{Fiziev}  {\small P. Fiziev, Gravitational Field of Massive Point Particle in General Relativity, gr-qc/0306088, ICTP preprint IC/2003/122. P.P. Fiziev, T.L. Bojadjiev, D.A. Georgieva, Novel Properties of Bound States of Klein-Gordon Equation in Gravitational Field ofMassive Point, gr-qc/0406036. P. Fiziev, S. Dimitrov, Point Electric Charge in General Relativity, hep-th/0406077. P. Fiziev, On the Solutions of Einstein Equations with Massive Point Source, gr-qc/0407088. }

\bibitem{Lichnerowicz}  {\small A. Lichnerowicz, "Theories Relativistes de la Gravitation et de l'Electromagnetisme," Masson, Paris (1955).}

\bibitem{Israel}  {\small W. Israel, "Singular hypersurfaces and thin shells in general relativity," Nuovo Cimento 44B, 1 (1966); and corrections in ibid. 48B, 463 (1966).}

\bibitem{Schwarzschild}  {\small K. Schwarzschild, Sitzungsber. Preus. Acad. Wiss. Phys. Math. Kl., 189 (1916). }

\bibitem{Misner}  {\small Misner, Thorne, Wheeler (1973). Gravitation. W H Freeman and Company. }

\bibitem{Misne}  {\small Misne E. Arnowitt, S. Deser, C. M. Misner, Phys. Rev. Lett.
D4, 375 (1960); Phys. Rev. 120 321 (1960).}

\end{thebibliography}
\end{document}